\documentstyle[12pt]{article}
\begin{document}

%%%  MORE DEFINITIONS
\def\D{\displaystyle}
\def\sn{\mathop{\hbox{\,\rm sn}}\nolimits} % Jacobi elliptic function sn
\def\dn{\mathop{\hbox{\,\rm dn}}\nolimits} % Jacobi elliptic function dn
\def\I{{\rm i}} % squart of -1
\def\rmd{{\rm d}}
\def\E#1{\exp(#1)} % exponential function
\def\hr{\hbox{I}\!\hbox{R}} % field of real numbers
\def\diag{\hbox{\rm diag}} % diagonal matrix
\def\re{\hbox{\rm Re}}  % real part of a complex number
\def\im{\hbox{\rm Im}}  % imaginary part of a complex number
\def\tr{\hbox{\rm tr}}
\def\Or{O}
\def\D{\displaystyle}
\def\dsum#1#2#3{\sum_{\scriptstyle #1\atop\scriptstyle #2}^{#3}}
\def\dsub#1#2{_{\scriptstyle #1\atop\scriptstyle #2}}

\def\zero{0}
\def\one{1}
\def\innerp#1#2#3{\if #3\zero\langle\Phi_{#1},\Phi_{#2}\rangle
 \else\if #3\one \langle\Phi_{#1},\Lambda\Phi_{#2}\rangle\else
 \langle\Phi_{#1},\Lambda^{#3}\Phi_{#2}\rangle\fi\fi}

\def\En#1#2{E_{#2}^{(#1)}}
\def\Ensq#1#2{E_{#2}^{(#1)2}}
\def\EEn#1#2{{\cal E}_{#2}^{(#1)}}
\def\TEEn#1#2{\widetilde {\cal E}_{#2}^{(#1)}}

%%% Unify the symbols in addresses and footnotes.
\newcounter{authorcount}
\setcounter{authorcount}{1}
\def\authorsymbol{\fnsymbol{authorcount}\addtocounter{authorcount}{1}}

% \fhl: half of \fl
% \fsl: one-fourth of \fl
\newdimen\mathindent
\mathindent=0pt
\newdimen\halfmathindent
\newdimen\smallmathindent
\halfmathindent=\mathindent
\smallmathindent=\mathindent
\divide\halfmathindent by 2
\divide\smallmathindent by 4
\advance\smallmathindent by \halfmathindent
\newcommand{\fl}{}
\newcommand{\fhl}{\hspace*{-\halfmathindent}}
\newcommand{\fsl}{\hspace*{-\smallmathindent}}

% \widetilde
\def\setboxz@h{\setbox\z@\hbox}
\def\wdz@{\wd\z@}
\def\boxz@{\box\z@}
\newif\ifmsbmloaded
\def\widetilde#1{\ifmsbmloaded\setboxz@h{$\m@th#1$}
 \ifdim\wdz@>\tw@ em\mathaccent"0\msbfam@5D{#1}\else
 \mathaccent"0365{#1}\fi\else\mathaccent"0365{#1}\fi}

% \allowmathbreak
\def\allowmathbreak{\relax\ifmmode\ifinner\allowbreak\else
  \nonmathaerr@\allowmathbreak\fi\else\nonmathberr@\allowmathbreak\fi}

% PROCLAIMS
\newtheorem{theorem}{\bf Theorem}
\newtheorem{lemma}{\bf Lemma}
\newtheorem{proposition}{\bf Proposition}
\newtheorem{corollary}{\bf Corollary}
\newtheorem{definition}{\bf Definition}
\newtheorem{example}{\sl Example}
\newtheorem{remark}{\sl Remark}
\def\demo{\vskip2pt\noindent{\sl Proof.}\hskip6pt\relax}
\def\enddemo{\vskip2pt\relax}

% REFERENCES
\def\voidtoken{} \def\refby{} \def\refpaper{} \def\refjour{}
\def\refvol{} \def\refpage{} \def\refpages{} \def\refyr{}
\def\refbook{} \def\refpubl{} \def\refpubladdr{}
\def\bibref{
  \global\def\refby{}   \global\def\refpaper{}
  \global\def\refjour{}   \global\def\refvol{}
  \global\def\refpage{}   \global\def\refpages{}
  \global\def\refyr{}   \global\def\refbook{}
  \global\def\refpubl{}   \global\def\refpubladdr{}
  }
\def\by#1{\global\def\refby{#1}}
\def\paper#1{\global\def\refpaper{#1}}
\def\jour#1{\global\def\refjour{#1}}
\def\vol#1{\global\def\refvol{#1}}
\def\page#1{\global\def\refpage{#1}}
\def\pages#1{\global\def\refpage{#1}}
\def\yr#1{\global\def\refyr{#1}}
\def\book#1{\global\def\refbook{#1}}
\def\publ#1{\global\def\refpubl{#1}}
\def\publaddr#1{\global\def\refpubladdr{#1}}
\def\endbibref{
\ifx\refby\voidtoken \else \refby\fi
\ifx\refyr\voidtoken \else \ \refyr\fi
\ifx\refpaper\voidtoken \else \ \refpaper\fi
\ifx\refjour\voidtoken \else \ {\it\refjour\/}\fi
\ifx\refbook\voidtoken \else \ {\it\refbook\/}\fi
\ifx\refpubl\voidtoken \else \ (\refpubl)\fi
\ifx\refpubladdr\voidtoken \else \ (\refpubladdr)\fi
\ifx\refvol\voidtoken \else \ {\bf\refvol}\fi
\ifx\refpage\voidtoken \else \ \refpage\fi\vskip2pt}

\title
{Finite-dimensional integrable systems associated with
Davey-Stewartson I equation}

\author
{\small Zixiang Zhou\\
\small Institute of Mathematics, Fudan University,
Shanghai 200433, China\\
\small E-mail: zxzhou@guomai.sh.cn\\
\small Wen-Xiu Ma\\
\small Department of Mathematics,
City University of Hong Kong, Kowloon, Hong Kong, China\\
\small E-mail: mawx@cityu.edu.hk\\
\small Ruguang Zhou\\
\small Department of Mathematics, Xuzhou Normal
University, Xuzhou 221009, China\\
\small E-mail: rgzhou@public.xz.js.cn
}
\date{}
\maketitle

\begin{abstract}
For the Davey-Stewartson I equation, which is an integrable
equation in 1+2 dimensions, we have already found its Lax pair in
1+1 dimensional form by nonlinear constraints. This paper deals
with the second nonlinearization of this 1+1 dimensional system to
get three 1+0 dimensional Hamiltonian systems with a constraint of
Neumann type. The full set of involutive conserved integrals is
obtained and their functional independence is proved. Therefore,
the Hamiltonian systems are completely integrable in Liouville
sense. A periodic solution of the Davey-Stewartson I equation is
obtained by solving these classical Hamiltonian systems as an
example.
\end{abstract}

\section{Introduction}

The Davey-Stewartson I (DSI) equation is a famous 1+2 dimensional
integrable equation describing the motion of water wave
\cite{bib:DS}. It has been discussed by various methods. Soliton
solutions were obtained by inverse scattering method
\cite{bib:Santini,bib:Pemp}, B\"acklund transformation
\cite{bib:Boiti}, binary Darboux transformation \cite{bib:MS},
nonlinearization method \cite{bib:Zhou2n,bib:Zhou2narb} etc.
Almost-periodic solutions were also obtained \cite{bib:JNS}.

In \cite{bib:Zhou2n,bib:Zhou2narb}, this 1+2 dimensional problem
was nonlinearized to an essentially 1+1 dimensional linear system
(\ref{eq:lp}) where all the differentials are separated. This
system is very useful to get localized soliton solutions by Darboux
transformation in 1+1 dimensions.

On the other hand, many 1+1 dimensional integrable systems can be
nonlinearized to 1+0 dimensional (or called finite-dimensional)
integrable systems \cite{bib:Cao0,bib:Ma,bib:Zeng,bib:ZRG}, and
the idea of nonlinearization was proposed by Cao \cite{bib:Cao0}.
Some 1+1 dimensional problems were completely solved and the
periodic or quasi-periodic solutions were obtained.

In \cite{bib:Caonew}, the KP equation, which is 1+2 dimensional, was
nonlinearized not only to 1+1 dimensional
\cite{bib:CL,bib:Kono}, but also to 1+0 dimensional integrable
Hamiltonian systems.

In the present paper, we show that the 1+1 dimensional system
obtained by nonlinearizing the Lax pair of the DSI equation can
also be nonlinearized to three 1+0 dimensional Hamiltonian systems.
We find a full set of involutive conserved integrals and prove
their functional independence. Therefore, these systems are
completely integrable in Liouville sense. As an example, when the
number of the eigenvalues is two, we solve the systems directly to
get a periodic solution of the DSI equation.

It is well known that the DSI equation has a Lax pair in 1+2
dimensions. In \cite{bib:Zhou2n,bib:Zhou2narb}, a new integrable
system was presented, which is essentially 1+1 dimensional, since
all the differentials are separated. This system can be written
explicitly as
\begin{equation}\fsl
   \begin{array}{l}
   \Phi_y=V\Phi=
   \left(\begin{array}{ccc}
   \I\lambda &u &\I f\\
   -\bar u &-\I\lambda &-\I g\\
   \I\bar f &-\I\bar g &0
   \end{array}\right)\Phi \qquad
   \Phi_x=U\Phi=\left(\begin{array}{ccc}
   \I\lambda &0 &\I f\\
   0 &\I\lambda &\I g\\
   \I\bar f &\I\bar g &0
   \end{array}\right)\Phi\\
   \Phi_t=W\Phi=\left(\begin{array}{ccc}
   -2\I\lambda^2+\I|u|^2+\I v_1 &-2u\lambda+\I u_y &-2\I f\lambda-2f_y\\
   2\bar u\lambda+\I\bar u_y &2\I\lambda^2-\I|u|^2-\I v_2
    &2\I g\lambda-2g_y\\
   -2\I\bar f\lambda+2\bar f_y &2\I\bar g\lambda+2\bar g_y
    &-2\I(|f|^2-|g|^2)
   \end{array}\right)\Phi.
\end{array} \label{eq:lp}
\end{equation}
Here $u$, $f$ and $g$ are complex functions, $v_1$ and $v_2$ are
real functions.

Its integrability conditions $\Phi_{xy}=\Phi_{yx}$,
$\Phi_{yt}=\Phi_{ty}$ and $\Phi_{xt}=\Phi_{tx}$
consist of the following three parts.

(1) DSI equation
\begin{equation}
   \begin{array}{l}
   -\I u_t=u_{xx}+u_{yy}+2|u|^2u+2(v_1+v_2)u\\
   v_{1,x}-v_{1,y}=v_{2,x}+v_{2,y}=-(|u|^2)_x.
\end{array} \label{eq:DS}
\end{equation}

(2) Standard Lax pair of DSI equation
\begin{equation}\fhl
   \begin{array}{l}
   F_y=\left(\begin{array}{ccc} 1 &0\\ 0 &-1\end{array}\right) F_x
    +\left(\begin{array}{ccc} 0 &u\\ -\bar u &0\end{array}\right) F\\
   F_t=2\I \left(\begin{array}{ccc} 1 &0\\ 0 &-1\end{array}\right) F_{xx}
    +2\I \left(\begin{array}{ccc} 0 &u\\ -\bar u &0\end{array}\right) F_x
    +\I\left(\begin{array}{ccc} |u|^2+2v_1 &u_x+u_y\\
    -\bar u_x+\bar u_y &-|u|^2-2v_2\end{array}\right) F
\end{array} \label{eq:orglp}
\end{equation}
where
\begin{equation}
   F=\left(\begin{array}{ccc} f\\g\end{array}\right).
\end{equation}

(3) Nonlinear constraint
\begin{equation}
   FF^*=\frac 12 \left(\begin{array}{ccc}
    v_1 &u_x\\ \bar u_x &v_2 \end{array}\right).
   \label{eq:1stconstr}
\end{equation}

Therefore, the nonlinear equations we will consider consist of
the DSI equation, its standard Lax pair in 1+2 dimensions and the
nonlinear constraint. As soon as this problem is solved, we get
the solution of the DSI equation, although only a part of the
solutions can be obtained by this method. In \cite{bib:Zhou2narb},
localized $(M,N)$ soliton solutions were obtained from the Darboux
transformation for (\ref{eq:lp}).

As many other 1+1 dimensional problems, here we want to find the
nonlinear constraint for (\ref{eq:lp}) and perform
nonlinearization again to obtain 1+0 dimensional Hamiltonian
systems.

\section{Nonlinearization and Hamiltonians}

In order to get a nonlinear constraint which is compatible with
all the $x$, $y$ and $t$-equations in (\ref{eq:lp}), we must
consider the $y$-equation first.

Notice that if $\Phi$ is a solution of (\ref{eq:lp}) for real
$\lambda$, then
\begin{equation}
   \bar\Phi_y=-V^T\bar\Phi\qquad
   \bar\Phi_x=-U^T\bar\Phi\qquad
   \bar\Phi_t=-W^T\bar\Phi.
\end{equation}
Suppose $\Phi=(\phi_1,\phi_2,\phi_3)^T$, then we may choose
$(\I \bar\phi_1,\I \bar\phi_2,\I \bar\phi_3)^T$ to be the
corresponding conjugate coordinates.

The first element of the Lenard sequence corresponding to
$w=(u,\,-\bar u,\,\I f,\,\I\bar f,\,\allowmathbreak
-\I g,\allowmathbreak\,-\I\bar g)$
is \cite{bib:Geng}
\begin{equation}
   G_0=(0,\, 0,\, \I\bar f,\, \I f,\, \I\bar g,\, \I g).
\end{equation}

By computing the variation of $\lambda$ \cite{bib:Ma}, we have
\begin{equation}
   \delta\lambda/\delta\omega=
   (\I\bar\phi_1\phi_2,\, \I\bar\phi_2\phi_1,\, \I\bar\phi_1\phi_3,\,
   \I\bar\phi_3\phi_1,\, \I\bar\phi_2\phi_3,\, \I\bar\phi_3\phi_2).
\end{equation}

Now take $N$ distinct eigenvalues $\lambda_j\in\hr$
($N\ge 2$). Suppose the corresponding solution of the Lax pair for
$\lambda=\lambda_j$ is $(\phi_{1j},\,\phi_{2j},\,\phi_{3j})^T$.
Then define
\begin{equation}
   \Lambda=\diag(\lambda_1,\cdots,\lambda_N)\qquad
   \Phi_j=(\phi_{j1},\cdots,\phi_{jN})^T\qquad
   (j=1,2,3).
\end{equation}
Each $\Phi_j$ is a column vector, and $\Lambda$ is a
constant real diagonal matrix.

We choose $(\phi_{1j},\phi_{2j},\phi_{3j},\I\bar\phi_{1j},
\I\bar\phi_{2j},\I\bar\phi_{3j})$ as the coordinates instead of
the real ones $\re(\phi_{jk})$ and $\im(\phi_{jk})$ $(1\le j\le
3,\,1\le k\le N)$. The standard symplectic form of $\hr^{6N}$ is
given by
\begin{equation}
   \omega=2\dsum{1\le j\le N}{1\le \alpha\le 3}{}
    \rmd\,\im(\phi_{j\alpha})\wedge\rmd\,\re(\phi_{j\alpha})
    =\dsum{1\le j\le N}{1\le \alpha\le 3}{}
    \I\,\rmd\bar\phi_{j\alpha}\wedge\rmd\phi_{j\alpha}.
   \label{eq:sympform}
\end{equation}

Denote $\langle V_1,V_2\rangle=V_1^*V_2$ for two column vectors
$V_1$ and $V_2$.

Define the nonlinear constraint
\begin{equation}
   G_0=\sum_j \delta\lambda_j/\delta\omega
   \label{eq:G=}
\end{equation}
which is
\begin{equation}
   \innerp 120=0\qquad
   \innerp 310=f\qquad
   \innerp 320=g.
   \label{eq:2ndconstr}
\end{equation}

By differentiating these constraints and using (\ref{eq:lp}), we have
\begin{equation}
   \begin{array}{l}
   \innerp 110_x=\innerp 220_x=\innerp 330_x=0\\
   \innerp 110_y=\innerp 220_y=\innerp 330_y=0\\
   \innerp 110_t=\innerp 220_t=\innerp 330_t=0
\end{array}
\end{equation}
\begin{equation}\fhl
   \begin{array}{l}
   f_x=\I\innerp 311+\I f(\innerp 330-\innerp 110)\\
   g_x=\I\innerp 321+\I g(\innerp 330-\innerp 220)\\
   f_y=\I\innerp 311+ug+\I f(\innerp 330-\innerp 110)\\
   g_y=-\I\innerp 321-\bar uf-\I g(\innerp 330-\innerp 220)
\end{array} \label{eq:fgxy}
\end{equation}
\begin{equation}\fhl
   u=\frac{2\I\innerp 211+2\I\innerp 230\innerp 310}{\innerp 110-\innerp 220}
    \label{eq:u}
\end{equation}
\begin{equation}\fhl
   u_x=2f\bar g \label{eq:ux}
\end{equation}
\begin{equation}\fhl
   \begin{array}{rl}
   u_y=q\equiv&\D\frac{2\I}{\innerp 110-\innerp 220}
   \big[2\I\innerp 212+2\I f\innerp 231+2\I\bar g\innerp 311\\
   &+\I f\bar g(2\innerp 330-\innerp 110-\innerp 220)\\
   &+u(|g|^2-|f|^2)+u(\innerp 221-\innerp 111)\big]
\end{array} \label{eq:uy}
\end{equation}
\begin{equation}\fhl
   v_1=2|f|^2\qquad v_2=2|g|^2. \label{eq:vexp}
\end{equation}
Therefore, the Lax pair (\ref{eq:lp}) becomes
\begin{eqnarray}
   &&\Phi_{1,x}=\I\Lambda\Phi_1+\I f\Phi_3 \qquad
   \Phi_{2,x}=\I\Lambda\Phi_2+\I g\Phi_3 \qquad
   \Phi_{3,x}=\I\bar f\Phi_1+\I\bar g\Phi_2
    \label{eq:newlpx}\\
   &&\Phi_{1,y}=\I\Lambda\Phi_1+u\Phi_2+\I f\Phi_3\qquad
   \Phi_{2,y}=-\bar u\Phi_1-\I\Lambda\Phi_2-\I g\Phi_3\nonumber\\
   &&\Phi_{3,y}=\I\bar f\Phi_1-\I\bar g\Phi_2
    \label{eq:newlpy}\\
   &&\Phi_{1,t}=(-2\I\Lambda^2+\I|u|^2+\I v_1)\Phi_1
    +(-2u\Lambda+\I u_y)\Phi_2+(-2\I f\Lambda-2f_y)\Phi_3
    \nonumber\\
   &&\Phi_{2,t}=(2\bar u\Lambda+\I\bar u_y)\Phi_1
    +(2\I\Lambda^2-\I|u|^2-\I v_2)\Phi_2
    +(2\I g\Lambda-2g_y)\Phi_3 \label{eq:newlpt}\\
   &&\Phi_{3,t}=(-2\I\bar f\Lambda+2\bar f_y)\Phi_1
    +(2\I\bar g\Lambda+2\bar g_y)\Phi_2
    -2\I(|f|^2-|g|^2)\Phi_3\nonumber
\end{eqnarray}
where $u$, $u_y$, $v_1$, $v_2$, $f$, $g$, $f_y$, $g_y$ are given
by (\ref{eq:2ndconstr})--(\ref{eq:vexp}) respectively.

Corresponding to the symplectic form (\ref{eq:sympform}),
the Poisson bracket of two functions $\xi$ and $\eta$ of
$\{\phi_{jk},\I\bar\phi_{jk}\}$ is given by
\begin{equation}
   \begin{array}{rl}
   \{\xi,\eta\}&\D=\frac12\sum_{j,k}\left(
    \frac{\partial\xi}{\partial\re(\phi_{jk})}
    \frac{\partial\eta}{\partial\im(\phi_{jk})}
    -\frac{\partial\xi}{\partial\im(\phi_{jk})}
    \frac{\partial\eta}{\partial\re(\phi_{jk})}\right)\\
   &\D=\frac 1\I\sum_{j,k}\left(
    \frac{\partial \xi}{\partial\phi_{jk}}
    \frac{\partial \eta}{\partial\bar\phi_{jk}}
    -\frac{\partial \xi}{\partial\bar\phi_{jk}}
    \frac{\partial \eta}{\partial\phi_{jk}}\right).
\end{array} \label{eq:Poisson}
\end{equation}
Hereafter, we always look $\phi_{jk}$ and $\bar\phi_{jk}$ as
independent variables in differentiations.

\begin{theorem}\label{thm:H}
The
Hamiltonians for (\ref{eq:newlpx}), (\ref{eq:newlpy}) and
(\ref{eq:newlpt}) are given by
\begin{eqnarray}
   \fsl H^x=&&-\innerp 111-\innerp 221-|\innerp 130|^2-|\innerp 230|^2
   \label{eq:Hx}\\
   \fsl H^y=&&-\innerp 111+\innerp 221-|\innerp 130|^2+|\innerp 230|^2
    \nonumber\\
   \fsl &&+\I u\innerp 120-\I\bar u\innerp 210 \label{eq:Hy}\\
   \fsl H^t=&&2\innerp 112-2\innerp 222
    +4\re(\innerp 130\innerp 311)\nonumber\\
   \fsl &&-4\re(\innerp 230\innerp 321)+2(\innerp 330-\innerp 110)
    \innerp 130\innerp 310\nonumber \\
   \fsl &&-2(\innerp 330-\innerp 220)\innerp 230\innerp 320
    \nonumber\\
   \fsl &&+\frac 4{\innerp 110-\innerp 220}
   \left|\innerp 211+\innerp 230\innerp 310\right|^2-2\re(q\innerp 120)
   \label{eq:Ht}
\end{eqnarray}
respectively, which satisfy
\begin{equation}
   \{H^x,\innerp 120\}=\{H^y,\innerp 120\}=\{H^t,\innerp 120\}=0
\end{equation}
Here $u$ is given by (\ref{eq:u}) and $q$ is given by
(\ref{eq:uy}).
\end{theorem}

The proof is obtained by direct computation.

Therefore, we obtain Hamiltonian systems
(\ref{eq:Hx})--(\ref{eq:Ht}) with the constraint of Neumann type
$\innerp 120=0$. Any solution of the Hamiltonian equations
\begin{equation}
   \begin{array}{l}
   \D\I\phi_{jk,x}=\frac{\partial H^x}{\partial\bar\phi_{jk}}\qquad
   \I\phi_{jk,y}=\frac{\partial H^y}{\partial\bar\phi_{jk}}\qquad
   \I\phi_{jk,t}=\frac{\partial H^t}{\partial\bar\phi_{jk}}\\
   \D-\I\bar\phi_{jk,x}=\frac{\partial H^x}{\partial\phi_{jk}}\qquad
   -\I\bar\phi_{jk,y}=\frac{\partial H^y}{\partial\phi_{jk}}\qquad
   -\I\bar\phi_{jk,t}=\frac{\partial H^t}{\partial\phi_{jk}}
\end{array} \label{eq:Heq}
\end{equation}
gives a solution of the DSI equation (\ref{eq:DS}), where $u$,
$v_1$, $v_2$ are given by (\ref{eq:u}) and (\ref{eq:vexp})
respectively.

\section{Integrability}

Now we consider the integrability of the Hamiltonian systems given
by Theorem~\ref{thm:H} on the submanifold
\begin{equation}\fl
   S=\{(\Phi_1,\Phi_2,\Phi_3,
   \I\bar\Phi_1,\I\bar\Phi_2,\I\bar\Phi_3)\in\hr^{6N}\,|\,
   \innerp 120=0,\,
   \innerp 110\ne\innerp 220\}.
   \label{eq:S}
\end{equation}
Here we still use $3N$ complex numbers and their complex conjugates
$(\Phi_1,\Phi_2,\allowmathbreak\Phi_3,\allowmathbreak
\I\bar\Phi_1,\I\bar\Phi_2,\I\bar\Phi_3)$ to represent a point in
$\hr^{6N}$. Clearly, $S$ has two connected components
characterized by $\innerp 110>\innerp 220$
and $\innerp 220>\innerp 110$ respectively.
Since $\innerp 110\ne\innerp 220$,
$0\not\in S$. $S$ is a $6N-2$ dimensional real analytic
manifold, on which the coordinates can be given by
$\phi_{31},\cdots\phi_{3N}$ (with their complex conjugates) and
$2N-1$ of $\phi_{11},\cdots,\phi_{1N},\phi_{21},\cdots,\phi_{2N}$
(with their complex conjugates) whenever the remaining one is non-zero.

Define
\begin{equation}
   \begin{array}{l}
   \D\gamma_1=\re\innerp 120=\frac 12(\innerp 120+\innerp 210)\\
   \D\gamma_2=\im\innerp 120=\frac 1{2\I}(\innerp 120-\innerp 210)
\end{array}\label{eq:gamma}
\end{equation}
then $S$ is defined by two real-valued functions
$\gamma_1=\gamma_2=0$. Since
\begin{equation}
   \{\gamma_1,\gamma_2\}
   =\frac 12(\innerp 110-\innerp 220)
\end{equation}
is never zero on $S$, the symplectic form (\ref{eq:sympform}) on
$\hr^{6N}$ naturally induces a (nondegenerate) symplectic form on
$S$. The corresponding Poisson bracket of two functions $\xi$,
$\eta$ on $S$ is still given by (\ref{eq:Poisson}) if they satisfy
$\{\xi,\gamma_j\}=0$, $\{\eta,\gamma_j\}=0$ $(j=1,2)$.

Hereafter, the Poisson bracket $\{\;\}$ always denotes the
standard Poisson bracket (\ref{eq:Poisson}) on $\hr^{6N}$.

Let
\begin{equation}
   L(\lambda)=\left(\begin{array}{ccc} 1 &&\\ &1&\\ &&0\end{array}\right)
   +\sum_{j=1}^N \frac 1{\lambda-\lambda_j}\left(\begin{array}{ccc}
   \bar\phi_{1j}\phi_{1j} &\bar\phi_{2j}\phi_{1j}
    &\bar\phi_{3j}\phi_{1j}\\
   \bar\phi_{1j}\phi_{2j} &\bar\phi_{2j}\phi_{2j}
    &\bar\phi_{3j}\phi_{2j}\\
   \bar\phi_{1j}\phi_{3j} &\bar\phi_{2j}\phi_{3j}
    &\bar\phi_{3j}\phi_{3j}
   \end{array}\right).
\end{equation}
Then we have

\begin{lemma}\label{lemma:Laxeq}
$L(\lambda)$ satisfies the Lax equations
\begin{equation}
   L_y=[V,L] \qquad L_x=[U,L] \qquad L_t=[W,L]
   \label{eq:Laxeq}
\end{equation}
if and only if the constraints (\ref{eq:2ndconstr}) hold.
\end{lemma}

\begin{demo}
Let $J=\diag(1,-1,0)$, $L_0=\diag(1,1,0)$,
$\phi_j=(\phi_{1j},\phi_{2j},\phi_{3j})^T$, then
\begin{equation}
   L(\lambda)=L_0+\sum_{j=1}^N\frac 1{\lambda-\lambda_j}\phi_j\phi_j^*.
\end{equation}
Since $V(\lambda_j)^*=-V(\lambda_j)$,
\begin{eqnarray}
   L_y(\lambda)&&=\sum_{j=1}^N\frac 1{\lambda-\lambda_j}
    [V(\lambda_j),\phi_j\phi_j^*]\nonumber\\
   &&=\sum_{j=1}^N\frac 1{\lambda-\lambda_j}[V(\lambda),\phi_j\phi_j^*]
    -\sum_{j=1}^N\frac 1{\lambda-\lambda_j}
    [V(\lambda)-V(\lambda_j),\phi_j\phi_j^*]\nonumber\\
   &&\D=[V(\lambda),L(\lambda)-L_0]-\I[J,\sum_{j=1}^N\phi_j\phi_j^*].
\end{eqnarray}
Hence $L_y(\lambda)=[V(\lambda),L(\lambda)]$ if and only if
\begin{equation}
   [L_0,V(\lambda)]=\I[J,\sum_{j=1}^N\phi_j\phi_j^*].
\end{equation}
Written in the components, this is exactly the constraints
(\ref{eq:2ndconstr}). This proves that the first equation of
(\ref{eq:Laxeq}) is equivalent to (\ref{eq:2ndconstr}). When
(\ref{eq:2ndconstr}) holds, the other two equations of
(\ref{eq:Laxeq}) are obtained similarly as the first one. The
lemma is proved.
\end{demo}

By Lemma~\ref{lemma:Laxeq}, $\tr L^k$ $(k\ge 1)$ are all conserved.
Expand $\tr L^k$ as a Laurent series
\begin{equation}
   \tr L^k=2+\sum_{j=0}^\infty
   \frac{\TEEn kj}{\lambda^{j+1}}
\end{equation}
which is convergent absolutely and uniformly as
$|\lambda|>\max_{1\le j\le N}|\lambda_j|$,
then all $\{\TEEn kj\}$ are conserved.

Moreover, we can show that any two of $\{\widetilde {\cal
E}_j^{(k)}\}$ commute with each other. This follows from the
following more general lemma.

\begin{lemma}\label{lemma:invol}
Suppose $\hr^{2nr}=\{(q_{11},\cdots,q_{1n},q_{21},\cdots,q_{2n},
\cdots,q_{r1},\cdots,q_{rn},p_{11},\cdots,p_{1n},\allowmathbreak
\cdots,\allowmathbreak
p_{r1},\allowmathbreak\cdots,\allowmathbreak p_{rn})\}$ is
equipped with the standard symplectic form
\begin{equation}
   \omega=\dsum{1\le j\le r}{1\le \alpha\le n}{}
   dp_{j\alpha}\wedge dq_{j\alpha}.
\end{equation}
Denote $q_j=(q_{j1},\cdots,q_{jn})^T$,
$p_j=(p_{j1},\cdots,p_{jn})^T$. Let $\lambda_1,\cdots,\lambda_n$
be $n$ distinct real numbers, $A$ be an $r\times r$ constant matrix,
\begin{equation}
   M(\lambda)=A+\sum_{\alpha=1}^n\frac 1{\lambda-\lambda_\alpha}
   \left(\begin{array}{cccc}
    p_{1\alpha}q_{1\alpha}
    & p_{2\alpha}q_{1\alpha} &\cdots
    & p_{r\alpha}q_{1\alpha}\\
    p_{1\alpha}q_{2\alpha}
    & p_{2\alpha}q_{2\alpha} &\cdots
    & p_{r\alpha}q_{2\alpha}\\
   \vdots &\vdots &\ddots &\vdots\\
    p_{1\alpha}q_{r\alpha}
    & p_{2\alpha}q_{r\alpha} &\cdots
    & p_{r\alpha}q_{r\alpha}\end{array}\right).
\end{equation}
Then for any two complex numbers $\lambda$, $\mu$ and any
positive integers $k$, $l$, $a$, $b$ with $1\le a,b\le r$,
\begin{equation}
   \begin{array}{l}
   \{\tr M^k(\lambda),\tr M^l(\mu)\}=0\\
   \{\tr M^k(\lambda),\langle p_b,q_a\rangle\}
    =-k[A,M^{k-1}(\lambda)]_{ab}.
\end{array}
\end{equation}
Here $\langle p_b,q_a\rangle=\sum_{\alpha=1}^n p_{b\alpha}q_{a\alpha}$.
\end{lemma}
\begin{demo}
The $j$-th row of $\D\frac{\partial
M(\lambda)}{\partial q_{j\alpha}}$ is
$\D\frac1{\lambda-\lambda_\alpha}
(p_{1\alpha},\cdots,p_{r\alpha})$ and the other rows are zero.
Similarly, the $j$-th column of $\D\frac{\partial
M(\lambda)}{\partial p_{j\alpha}}$ is $\D\frac
1{\lambda-\lambda_\alpha}
(q_{1\alpha},\cdots,q_{r\alpha})^T$ and the other columns are
zero. Hence
\begin{equation}\fsl
   \begin{array}{rl}
   &\D\frac 1{kl}\left\{\tr M^k(\lambda),\tr M^l(\mu)\right\}\\
   =&\D\frac 1{kl}\sum_{j,\alpha}\Bigg(
    \frac{\partial}{\partial q_{j\alpha}}\big(\tr M^k(\lambda)\big)
    \frac{\partial}{\partial p_{j\alpha}}\big(\tr M^l(\mu)\big)
    -\frac{\partial}{\partial q_{j\alpha}}\big(\tr M^l(\mu)\big)
    \frac{\partial}{\partial p_{j\alpha}}\big(\tr M^k(\lambda)\big)\Bigg)\\
   =&\D\sum_{j,\alpha}\Bigg(
    \tr\Big(M^{k-1}(\lambda)\frac{\partial M(\lambda)}{\partial q_{j\alpha}}
     \Big)
    \tr\Big(M^{l-1}(\mu)\frac{\partial M(\mu)}{\partial p_{j\alpha}}\Big)\\
   &\D-\tr\Big(M^{l-1}(\mu)\frac{\partial M(\mu)}{\partial q_{j\alpha}}\Big)
    \tr\Big(M^{k-1}(\lambda)\frac{\partial M(\lambda)}{\partial p_{j\alpha}}
     \Big)\Bigg)\\
   =&\D\sum_{a,b,j,\alpha}\Bigg(
    \frac 1{\lambda-\lambda_\alpha}(M^{k-1}(\lambda))_{aj}p_{a\alpha}\cdot
    \frac 1{\mu-\lambda_\alpha}(M^{l-1}(\mu))_{jb}q_{b\alpha}\\
    &\D-\frac 1{\mu-\lambda_\alpha}(M^{l-1}(\mu))_{aj}p_{a\alpha}\cdot
    \frac 1{\lambda-\lambda_\alpha}(M^{k-1}(\lambda))_{jb}q_{b\alpha}\Bigg)\\
   =&\D\sum_{a,b,\alpha}\frac 1{\lambda-\lambda_\alpha}
     \frac 1{\mu-\lambda_\alpha}
    p_{a\alpha}q_{b\alpha}[M^{k-1}(\lambda),M^{l-1}(\mu)]_{ab}\\
   =&\D\sum_{a,b,\alpha}\frac 1{\mu-\lambda}\Bigg(
    \frac 1{\lambda-\lambda_\alpha}-\frac 1{\mu-\lambda_\alpha}\Bigg)
    p_{a\alpha}q_{b\alpha}[M^{k-1}(\lambda),M^{l-1}(\mu)]_{ab}\\
   =&\D\frac 1{\mu-\lambda}\tr\Bigg((M(\lambda)-M(\mu))
    [M^{k-1}(\lambda),M^{l-1}(\mu)]\Bigg)=0.
\end{array}
\end{equation}
This proves the first part. The second part is proved as follows.
\begin{equation}\fhl
   \begin{array}{rl}
   &\D\frac 1{k}\left\{\tr M^k(\lambda),\langle p_b,q_a\rangle\right\}\\
   =&\D\frac 1{k}\sum_{\alpha}\Bigg(
    \frac{\partial}{\partial q_{b\alpha}}\big(\tr M^k(\lambda)\big)
    q_{a\alpha}
    -\frac{\partial}{\partial p_{a\alpha}}\big(\tr M^k(\lambda)\big)
    p_{b\alpha}\Bigg)\\
   =&\D\sum_{l,\alpha}\Bigg(
    \frac1{\lambda-\lambda_\alpha}(M^{k-1}(\lambda))_{lb}
    p_{l\alpha}q_{a\alpha}
    -\frac1{\lambda-\lambda_\alpha}(M^{k-1}(\lambda))_{al}
    q_{l\alpha}p_{b\alpha}
    \Bigg)\\
   =&\D\bigg((M(\lambda)-A)M^{k-1}(\lambda)
    -M^{k-1}(\lambda)(M(\lambda)-A)\bigg)_{ab}\\
   =&\D-[A,M^{k-1}(\lambda)]_{ab}.
\end{array}
\end{equation}
The lemma is proved.
\end{demo}

From this lemma, we know that the set $\{\widetilde {\cal
E}_j^{(k)}\}$ is in involution.

\begin{remark}
For the three-wave equation, \cite{bib:Geng} wrote down such
$\{\TEEn kj\}$. It was proved that they were in involution and
$3N$ of them were independent. However, for the three-wave
equation, the first term of $L$ is
$\diag(\beta_1,\beta_2,\beta_3)$ ($\beta_1$, $\beta_2$, $\beta_3$
are distinct) rather than $\diag(1,1,0)$ here. In the
case $\beta_1=1$, $\beta_2=1$, $\beta_3=0$, those $\{\TEEn kj\}$
are obviously still in involution. Hence we can also use the
result of \cite{bib:Geng} to get the involution. But we prove it
more directly and easily here. On the other hand, with the
constraint $\innerp 120=0$, the independence of $\{\TEEn kj\}$
completely changes. Hence we should prove the independence in
this constrained case.
\end{remark}

For real $\lambda$, suppose the eigenvalues of the Hermitian
matrix $L(\lambda)$ are $\nu_1$, $\nu_2$ and $\nu_3$, then $\tr
L^k=\sum_{j=1}^3\nu_j^k$, while
\begin{equation}
   \begin{array}{l}
    \D\det(\mu-L(\lambda))=\prod_{j=1}^3(\mu-\nu_j)\\
   =\mu^3-(\nu_1+\nu_2+\nu_3)\mu^2
   +(\nu_1\nu_2+\nu_2\nu_3+\nu_3\nu_1)\mu-\nu_1\nu_2\nu_3
\end{array}
\end{equation}
for any $\mu$. Hence $\tr L^k$ can be expressed by the
coefficients of $\mu$ in $\det(\mu-L(\lambda))$.

Suppose $L(\lambda)=(L_{jk})_{1\le j,k\le 3}$, then
\begin{equation}
   \begin{array}{l}
   \nu_1+\nu_2+\nu_3=\tr L\qquad\nu_1\nu_2\nu_3=\det L\\
   \D\nu_1\nu_2+\nu_2\nu_3+\nu_3\nu_1=\sum_{1\le j<k\le 3}\left|
   \begin{array}{cc}
    L_{jj} &L_{jk}\\ L_{kj} &L_{kk} \end{array}\right|.
\end{array} \end{equation}
Define
\begin{equation}
   \begin{array}{l}
   \D\tr L=\sum_{k=-1}^\infty\frac{\EEn 0k}{\lambda^{k+1}}
    \qquad
   \D\det L=\sum_{k=-1}^\infty\frac{\EEn 2k}{\lambda^{k+1}}\\
   \D\sum_{1\le j<k\le 3}\left|\begin{array}{cc}
    L_{jj} &L_{jk}\\ L_{kj} &L_{kk} \end{array}\right|
    =\sum_{k=-1}^\infty\frac{\EEn 1k}{\lambda^{k+1}}.
\end{array}
\end{equation}
These $\{\EEn jk\}$ differ from the $\{F_k^{(j)}\}$ in
\cite{bib:Geng} with a multiple $i$.

To simplify the expressions of $\EEn jk$'s, we define the
following simpler but equivalent conserved integrals $\En jk$'s,
which are nondegenerate linear combinations of $\EEn jk$'s. This
also makes the expressions (\ref{eq:Hxexp})--(\ref{eq:Htexp}) of
the Hamiltonians and the proof of Theorem~\ref{thm:indep} simpler.
For $m\ge 0$, define
\begin{equation}\fhl
   \begin{array}{rl}
   \En1m&=2\EEn 0m-\EEn 1m\\
   &=\innerp 11m+\innerp 22m\\
   &\D\quad-\sum_{1\le i<j\le 3}\sum_{l=1}^m\left|\begin{array}{cc}
    \innerp ii{l-1} &\innerp ji{m-l}\\
    \innerp ij{l-1} &\innerp jj{m-l} \end{array}\right|\\
\end{array}\label{eq:Em1}
\end{equation}
\begin{equation}\fhl
   \begin{array}{rl}
   \En2m&=-\EEn 2{m+1}+\EEn 1{m+1}-\EEn 0{m+1}\\
   &\D=\sum_{l=0}^{m}\left|\begin{array}{cc}
    \innerp 11l &\innerp 21{m-l}\\
    \innerp 12l &\innerp 22{m-l}\end{array}\right|\\
   &\D\quad-\dsum{i+j+k=m-1}{i,j,k\ge 0}{}\left|\begin{array}{ccc}
    \innerp 11i &\innerp 21j &\innerp 31k\\
   \innerp 12i &\innerp 22j &\innerp 32k\\
   \innerp 13i &\innerp 23j &\innerp 33k\end{array}\right|
\end{array}\label{eq:Em2}
\end{equation}
\begin{equation}\fhl
   \begin{array}{rl}
   \En3m&=\EEn 1m-\EEn 0m
    =\innerp 33m\\
    &\D\quad+\sum_{1\le i<j\le 3}\sum_{l=1}^m\left|\begin{array}{cc}
    \innerp ii{l-1} &\innerp ji{m-l}\\
    \innerp ij{l-1} &\innerp jj{m-l} \end{array}\right|.
\end{array}\label{eq:Em3}
\end{equation}
The above sums are zero if the upper bound is smaller than
the lower bound. According to Lemma~\ref{lemma:invol}, we have

\begin{theorem}\label{thm:involution}
$\{\En jm,\En kn\}=0$ and $\{\En jm,\innerp 120\}=0$ for any
$j,k=1,2,3$ and $m,n\ge 0$. Therefore, $\{\En jm\}$ are in
involution on $S$.
\end{theorem}

Define
\begin{equation}\fhl
   \Omega_1=\innerp 110\qquad
   \Omega_2=\innerp 220\qquad
   \Omega_3=\innerp 330.
\end{equation}
By (\ref{eq:Em1})--(\ref{eq:Em3}),
\begin{equation}
   \En10=\Omega_1+\Omega_2\qquad
   \En20=\Omega_1\Omega_2-|\innerp120|^2\qquad
   \En30=\Omega_3.
   \label{eq:Omegarel}
\end{equation}
Hence $\Omega_1$, $\Omega_2$, $\Omega_3$ are expressed by $\En
jk$'s and $|\innerp 120|^2$.

The Hamiltonians in Theorem~\ref{thm:H} can be expressed as
\begin{eqnarray}
   &&H^x=-\En11-\En10\En30-\En20 \label{eq:Hxexp}\\
   &&H^y=\frac1{\Omega_1-\Omega_2}(-\En10\En11
    +2\En21-\En10\En20-\Ensq10\En30+2\En20\En30)
    \label{eq:Hyexp}\\
   &&H^t=\frac1{\Omega_1-\Omega_2}\bigg(
    2\En10\En12-4\En22+2(\Ensq10-2\En20)(\En11+\En31)\nonumber\\
   &&\qquad+(\Ensq10-4\En20)H^x+(\En10-2\En30)(\Omega_1-\Omega_2)H^y
   +(H^x)^2-(H^y)^2\nonumber\\
   &&\qquad-(u\innerp 120-\bar u\innerp 210)^2
    +4(|\innerp 130|^2+|\innerp 230|^2)
     |\innerp 120|^2\bigg).\label{eq:Htexp}
\end{eqnarray}
Each Hamiltonian is a function of $\En jk$'s and
$|\innerp 120|^2$, $\innerp 120^2$, $\innerp 210^2$, which is
smooth near $S$.

Let $\gamma_1$ and $\gamma_2$ be two real functions in (\ref{eq:gamma})
defining $S$. For any two functions $H_1$ and $H_2$ on $\hr^{6N}$
with $\{H_j,\gamma_k\}=0$ $(j,k=1,2)$, define $H_1\doteq 
H_2$ if $H_1=H_2$ and $\nabla H_1=\nabla H_2$ on $S$. In this
case, $H_1$ and $H_2$ have the same Hamiltonian vector field
which is tangent to $S$. Hence they give the same Hamiltonian
systems on $S$. Moreover, for any function $F$ on $\hr^{6N}$ with
$\{F,\gamma_k\}=0$ $(k=1,2)$, $\{H_1,F\}=\{H_2,F\}$ holds.

From (\ref{eq:Omegarel}),
\begin{equation}\fhl
   \Omega_1-\Omega_2=\sigma^{-1}\sqrt{\Ensq10-4(\En20+|\innerp 120|^2)}
\end{equation}
where $\sigma=\pm1$ depends on the connected component of $S$. Let
\begin{equation}\fhl
   \begin{array}{l}
   \Delta=\sqrt{\Ensq10-4\En20}\\
   K^y=(\Omega_1-\Omega_2)H^y=-\En10\En11+2\En21-\En10\En20
    -\Ensq10\En30+2\En20\En30.
\end{array}
\end{equation}
For $\gamma_1$ and $\gamma_2$, only the quadratic terms appear in the
expressions of $H^y$ and $H^t$. Hence we have
\begin{eqnarray}
   &&H^y\doteq\frac\sigma\Delta K^y \label{eq:Hyexp2}\\
   &&H^t\doteq\frac\sigma\Delta\bigg(
    2\En10\En12-4\En22+2(\Ensq10-2\En20)(\En11+\En31)\nonumber\\
   &&\qquad+(\Ensq10-4\En20)H^x+(\En10-2\En30)K^y
   +(H^x)^2-\frac 1{\Delta^2}(K^y)^2\bigg).\label{eq:Htexp2}
\end{eqnarray}
By (\ref{eq:Hxexp}), $H^x$ depends on $\En jk$'s only. On the
other hand, the right hand sides of (\ref{eq:Hyexp2}) and
(\ref{eq:Htexp2}) also depend on $\En jk$'s only. Therefore, we
get the following result from Theorem~\ref{thm:H} and
Theorem~\ref{thm:involution}. 

\begin{theorem}
Three Hamiltonians $H^x$, $H^y$, $H^t$ defined by
Theorem~\ref{thm:H} commute with each other:
\begin{equation}
   \{H^x,H^y\}=\{H^x,H^t\}=\{H^y,H^t\}=0
\end{equation}
and they satisfy
\begin{equation}
   \{H^x,\innerp 120\}=\{H^y,\innerp 120\}=
   \{H^t,\innerp 120\}=0.
\end{equation}
Moreover, each $\En jm$ is conserved under the Hamiltonian
flows given by $H^x$, $H^y$, $H^t$ respectively.
\end{theorem}

Next we shall prove the integrability of these Hamiltonian systems.
That is
\begin{theorem}\label{thm:indep}
$3N-1$ real-valued functions $\{\En jm\}$ $(j=1,3;\,0\le m\le
N-1)$ and $\{\En 2m\}$ $(0\le m\le N-2)$ are functionally
independent in a dense open subset of $S$.
\end{theorem}
\begin{demo}
Let
\begin{equation}\fsl
   \begin{array}{l}
   \widetilde S_1=\{(\Phi_1,\Phi_2,\Phi_3,
   \I\bar\Phi_1,\I\bar\Phi_2,\I\bar\Phi_3)\in S\,|\,
   \phi_{1,N}\ne 0,\,\bar\phi_{1,N}\ne 0,\,
   \innerp 110>\innerp 220\}\\
   \widetilde S_2=\{(\Phi_1,\Phi_2,\Phi_3,
   \I\bar\Phi_1,\I\bar\Phi_2,\I\bar\Phi_3)\in S\,|\,
   \phi_{1,N}\ne 0,\,\bar\phi_{1,N}\ne 0,\,
   \innerp 110<\innerp 220\}\\
   \widetilde S=\widetilde S_1\cup\widetilde S_2
\end{array}
\end{equation}
then $\widetilde S$ is a dense open subset of $S$. Similar to
$S$, $\widetilde S$ has also two connected components, which are
$\widetilde S_1$ and $\widetilde S_2$. In $\widetilde S$, we can
solve $\phi_{2,N}$, $\bar\phi_{2,N}$ from the constraint
$\innerp 120=0$ as
\begin{equation}
   \phi_{2,N}=-\sum_{j=1}^{N-1}\frac{\bar\phi_{1j}}{\bar\phi_{1N}}\phi_{2j}
   \qquad
   \bar\phi_{2,N}=-\sum_{j=1}^{N-1}\frac{\phi_{1j}}{\phi_{1N}}\bar\phi_{2j}.
\end{equation}
Hence $\widetilde S_1$ has global coordinates
\begin{equation}\fhl
   \Theta=\{\phi_{1j},\I\bar\phi_{1j}\,(1\le j\le N);\,
   \phi_{2j},\I\bar\phi_{2j}\,(1\le j\le N-1);\,
   \phi_{3j},\I\bar\phi_{3j}\,(1\le j\le N)\}.
\end{equation}

Let $P_0\in \widetilde S_1$ be given by $\Phi_1=(1,1,\cdots,1)^T$,
$\Phi_2=\epsilon(1,1,\cdots,1,-N+1)^T$,
$\Phi_3=\epsilon(1,1,\cdots,1)^T$ where $\epsilon$ is a small
real constant. Here $\Phi_3$ is chosen to be parallel with $\Phi_1$ so
that the following computation will be simplified.
Since $\phi_{2N}$ and $\bar\phi_{2N}$ are
functions of the variables in $\Theta$, we have, at $P_0$,
\begin{equation}\fl
   \begin{array}{l}
   \D\frac{\partial \En1m}{\partial\bar\phi_{1j}}
    =\lambda_j^m+\Or(\epsilon)\qquad
    \frac{\partial\En1m}{\partial\bar\phi_{2j}}=\Or(\epsilon)\qquad
    \frac{\partial\En1m}{\partial\bar\phi_{3j}}=\Or(\epsilon)\qquad\\
   \D\frac{\partial\En2m}{\partial\bar\phi_{1j}}=\Or(\epsilon^2)
    \qquad\frac{\partial\En2m}{\partial\bar\phi_{3j}}=\Or(\epsilon^3)\\
   \D\frac{\partial\En2m}{\partial\bar\phi_{2j}}=
   \sum_{l=0}^m \left|\begin{array}{cc}
    \innerp 11l &\lambda_j^{m-l}\phi_{1j}\\
    \innerp 12l &\lambda_j^{m-l}\phi_{2j}
   \end{array}\right|
   -\frac{\phi_{1j}}{\phi_{1N}}\sum_{l=0}^m\left|\begin{array}{cc}
    \innerp 11l &\lambda_N^{m-l}\phi_{1N}\\
    \innerp 12l &\lambda_N^{m-l}\phi_{2N}
   \end{array}\right|
    +\Or(\epsilon^3)\\
   \D\qquad=\epsilon\sum_{l=0}^m\left|\begin{array}{cc}
    \lambda_1^l+\cdots+\lambda_N^l &\lambda_j^{m-l}-\lambda_N^{m-l}\\
    \lambda_1^l+\cdots+\lambda_{N-1}^l-(N-1)\lambda_N^l &\lambda_j^{m-l}
    +(N-1)\lambda_N^{m-l}
   \end{array}\right|+\Or(\epsilon^3)\\
   \D\qquad=N\epsilon\sum_{l=0}^m\lambda_N^{m-l}(\lambda_1^l+\cdots
    +\lambda_{N-1}^l+\lambda_j^l)+\Or(\epsilon^3)\\
   \D\frac{\partial\En3m}{\partial\bar\phi_{1j}}=\Or(\epsilon^2)\qquad
   \frac{\partial\En3m}{\partial\bar\phi_{2j}}=\Or(\epsilon)\qquad
   \frac{\partial\En3m}{\partial\bar\phi_{3j}}=\epsilon\lambda_j^m
   +\Or(\epsilon^3).
\end{array}
\end{equation}
Here the subscript $j$ is taken from $1$ to $N$ for $\phi_{1j}$,
$\phi_{3j}$, and from $1$ to $N-1$ for $\phi_{2j}$.
It can be checked that
\begin{equation}\fhl
   \det\left(\sum_{l=0}^m \lambda_N^{m-l}(\lambda_1^l+\cdots
    +\lambda_{N-1}^l+\lambda_j^l)\right)\dsub{0\le m\le N-2}{1\le j\le N-1}
   =N\prod_{1\le j<k\le N-1}(\lambda_j-\lambda_k).
\end{equation}
Hence the Jacobian determinant
\begin{equation}\fhl
   \begin{array}{rl}
   J&\D\equiv\frac{\partial(\En10,\cdots,\En1{N-1},
   \En20,\cdots,\En2{N-2},
   \En30,\cdots,\En3{N-1})}
   {\partial(\bar\phi_{11},\cdots,\bar\phi_{1,N},
   \bar\phi_{21},\cdots,\bar\phi_{2,N-1},
   \bar\phi_{31},\cdots,\bar\phi_{3,N})}\\
   &\D=N^N\epsilon^{2N-1}\left(\prod_{1\le j<k\le N}
    (\lambda_j-\lambda_k)\right)^2
    \prod_{1\le j<k\le N-1}(\lambda_j-\lambda_k)+\Or(\epsilon^{2N}).
\end{array}
\end{equation}
When $\epsilon$ is small enough and $\epsilon\ne 0$, $J$ is
non-zero near the point $P_0$. Since $J$ is a rational function
of
\begin{equation}\fhl
   \{\phi_{1j},\I\bar\phi_{1j}\,(1\le j\le N);\,
   \phi_{2j},\I\bar\phi_{2j}\,(1\le j\le N-1);\,
   \phi_{3j},\I\bar\phi_{3j}\,(1\le j\le N)\}
\end{equation}
$J=0$ identically if $J$ is zero in an open subset of
$\widetilde S_1$. Therefore, $J$ is non-zero in a
dense open subset of $\widetilde S_1$. Similarly, $J$ is non-zero
in a dense open subset of $\widetilde S_2$. Since all $\En jm$
are real-valued functions, the Jacobian matrix of
$(\En10,\cdots,\En1{N-1}, \En20,\cdots,\En2{N-2},
\En30,\cdots,\En3{N-1})$ with respect to the real coordinates
\begin{equation}\fhl
   \begin{array}{l}
   \re(\phi_{11}),\cdots,\re(\phi_{1,N}),
   \re(\phi_{21}),\cdots,\re(\phi_{2,N-1}),
   \re(\phi_{31}),\cdots,\re(\phi_{3,N}),\\
   \im(\phi_{11}),\cdots,\im(\phi_{1,N}),
   \im(\phi_{21}),\cdots,\im(\phi_{2,N-1}),
   \im(\phi_{31}),\cdots,\im(\phi_{3,N})
\end{array}
\end{equation}
is of full rank $3N-1$. The theorem is proved.
\end{demo}

\begin{theorem}
The Hamiltonian systems given by Theorem~\ref{thm:H} are completely
integrable on $S$ in Liouville sense.
\end{theorem}
\begin{demo}
We have proved: (1) $\{\En jm\}$ $(j=1,2,3;m=0,1,2,\cdots)$
are in involution on $S$ (Theorem~\ref{thm:involution}).
(2) $\{\En jm\}$ $(j=1,3;\,0\le m\le N-1)$ and $\{\En2m\}$
$(0\le m\le N-2)$ are functionally independent in a dense open
subset of $S$ (Theorem~\ref{thm:indep}). It remains to prove that
the Hamiltonian vector fields of all $\{\En jm\}$ are complete.
This follows from the compactness of each level set, which is
a closed subset of the compact set
\begin{equation}
   \{(\Phi_1,\Phi_2,\Phi_3,
   \I\bar\Phi_1,\I\bar\Phi_2,\I\bar\Phi_3)\in S\,|\,
   \innerp jj0=\Omega_{j0}, j=1,2,3\}
\end{equation}
where $\Omega_{j0}$ $(j=1,2,3)$ are constants. Therefore, the
Hamiltonian systems given by Theorem~\ref{thm:H} are completely
integrable \cite{bib:Cushman}.
\end{demo}

\section{Example: An explicit solution for $N=2$}

Now suppose $N=2$,
$\Lambda=\D \left(\begin{array}{ccc}\lambda
&0\\0&\mu\end{array}\right)$, $\D
\Phi_j=\left(\begin{array}{ccc}\phi_j\\\psi_j\end{array}\right)$.
Let $R_1$, $R_2$, $R_3$, $G$ and $K$ be defined by
\begin{equation}
   \begin{array}{l}
   R_j^2=|\phi_j|^2+|\psi_j|^2\quad(j=1,2,3)\\
   G=|\phi_1|^2+|\phi_2|^2+|\phi_3|^2-R_2^2\\
   K=(\innerp 111
    +\innerp 130\innerp 310)/R_1^2
\end{array} \label{eq:egconserve}
\end{equation}
then from the above list of conserved integrals, we know that
$R_1$, $R_2$, $R_3$, $G$ and $K$ are all constants.
Moreover,
\begin{equation}
   \innerp 221
    +\innerp 230\innerp 320
    =R_2^2(R_3^2+\lambda+\mu-K).
\end{equation}

Let
\begin{equation}
   \phi_j=R_j\cos\theta_j\E{\I\alpha_j}\qquad
   \psi_j=R_j\sin\theta_j\E{\I\beta_j}
   \label{eq:egphipsi}
\end{equation}
then the constraint $\innerp 210=0$ leads to
\begin{equation}
   \theta_2=\sigma\theta_1+\pi/2+l\pi\qquad
   \E{\I(\beta_2-\beta_1-\alpha_2+\alpha_1)}=\sigma
\end{equation}
where $l$ is an integer and $\sigma=\pm 1$. Notice that
(\ref{eq:egphipsi}) is invariant under the transformation
$\theta_1\to-\theta_1$, $\beta_1\to\beta_1+\pi$. Hence we can
always choose $\sigma=1$.

Let $\delta=\beta_1-\beta_3-\alpha_1+\alpha_3$ and
$\rho=\cos^2\theta_1$. Substituting (\ref{eq:egphipsi}) into
the second equation of (\ref{eq:egconserve}), we obtain
\begin{equation}
   R_3^2\cos^2\theta_3=G-(R_1^2-R_2^2)\rho.
\end{equation}
The third equation of (\ref{eq:egconserve}) gives
\begin{equation}
   \begin{array}{l}
    \lambda\cos^2\theta_1+\mu\sin^2\theta_1
   +R_3^2(\cos^2\theta_1\cos^2\theta_3+\sin^2\theta_1\sin^2\theta_3)\\
   +2R_3^2\cos\theta_1\sin\theta_1\cos\theta_3
    \sin\theta_3\cos\delta=K
\end{array} \label{eq:cosdelta}
\end{equation}
from which $\delta$ can be solved as a function of $\rho$.

From the equation (\ref{eq:newlpx}), we have
\begin{equation}
   \begin{array}{l}
   \theta_{1,x}=R_3^2\cos\theta_3\sin\theta_3\sin\delta\\
   \theta_{3,x}=(R_2^2-R_1^2)\cos\theta_1\sin\theta_1\sin\delta\\
   \alpha_{1,x}=\lambda+R_3^2\cos^2\theta_3
    +R_3^2\cos\theta_3\sin\theta_3\tan\theta_1\cos\delta\\
   \alpha_{2,x}=\lambda+R_3^2\cos^2\theta_3
    -R_3^2\cos\theta_3\sin\theta_3\cot\theta_1\cos\delta\\
   \alpha_{3,x}=R_1^2\cos^2\theta_1+R_2^2\sin^2\theta_1
    +(R_1^2-R_2^2)\cos\theta_1\sin\theta_1\tan\theta_3\cos\delta\\
   \beta_{1,x}=\mu+R_3^2\sin^2\theta_3
    +R_3^2\cos\theta_3\sin\theta_3\cot\theta_1\cos\delta\\
   \beta_{2,x}=\mu+R_3^2\sin^2\theta_3
    -R_3^2\cos\theta_3\sin\theta_3\tan\theta_1\cos\delta\\
   \beta_{3,x}=R_1^2\sin^2\theta_1+R_2^2\cos^2\theta_1
    +(R_1^2-R_2^2)\cos\theta_1\sin\theta_1\cot\theta_3\cos\delta.
   \end{array}
\end{equation}

The first equation of the above system leads to
\begin{equation}
   \rho_x=-2R_3^2\cos\theta_1\sin\theta_1\cos\theta_3\sin\theta_3\sin\delta.
\end{equation}
Solving $\cos\delta$ from (\ref{eq:cosdelta}), we get the
equation of $\rho$:
\begin{equation}
   \rho_x=-\sqrt{P(\rho)}
\end{equation}
where
\begin{equation}\fl
   \begin{array}{rl}
   P(\rho)=&4b(\lambda-\mu)\rho^3
   +\big(4(\mu-\lambda)G+4b(\mu-K)-(\lambda-\mu+b-a)^2\big)\rho^2\\
   &+\big(4(K-\mu)G+2(\lambda-\mu+b-a)(K+G-\mu-a)\big)\rho-(K+G-\mu-a)^2
\end{array}
   \label{eq:P}
\end{equation}
which is a cubic polynomial,
\begin{equation}
   a=R_3^2 \qquad b=R_1^2-R_2^2.
\end{equation}
Suppose $b>0$, $\lambda>\mu$ and $P$ has three different real roots
$\rho_1<\rho_2<\rho_3$. Moreover, suppose
\begin{equation}
   \begin{array}{l}
   K+G-\mu-a\ne 0 \qquad
   K-G-\lambda+b\ne 0 \\
   \max\bigg(0,\frac{G-a}{b}\bigg)<\min\bigg(1,\frac{G}{b}\bigg)
\end{array} \label{eq:KGcond}
\end{equation}
Then the solution $\rho$ can be expressed by elliptic functions
of $x$. Let $\rho=\rho_1+(\rho_2-\rho_1)\omega^2$, then
\begin{equation}
   \omega_x=\pm p\sqrt{(1-\omega^2)(1-k^2\omega^2)}
\end{equation}
where
\begin{equation}
   k=\sqrt{(\rho_2-\rho_1)/(\rho_3-\rho_1)} \qquad
   p=\sqrt{b(\lambda-\mu)(\rho_3-\rho_1)}. \label{eq:k-p}
\end{equation}
Hence $\omega=\pm\sn(p(x-\widetilde x_0))$,
\begin{equation}
   \rho=\rho_1+(\rho_2-\rho_1)\sn^2(p(x-\widetilde x_0))
\end{equation}
where $\widetilde x_0$ is independent of $x$, but may depend on
$y$ and $t$. $\rho$ is a periodic function of $x$.
\begin{remark}
Since $\rho_j$ is a root of $P$, (\ref{eq:P}) leads to
\begin{eqnarray}\fhl
   \begin{array}{l}
   4\rho_j(1-\rho_j)(G-b\rho_j)(a-G+b\rho_j)\\
   =\bigg(K-\lambda\rho_j-\mu(1-\rho_j)-\rho_j(G-b\rho_j)
   -(1-\rho_j)(a-G+b\rho_j)\bigg)^2\ge 0.
\end{array}
\end{eqnarray}
(This is actually equivalent to (\ref{eq:cosdelta})). Hence,
$\max(0,\frac{G-a}{b})\le\rho_1<\rho_2\le\min(1,\frac{G}{b})$
holds if there is a solution locally,
since $0\le\cos^2\theta_1\le 1$ and $0\le\cos^2\theta_3\le 1$
should be satisfied. This also guarantees that the solution is
global because $\rho_1\le\rho\le\rho_2$.  Moreover, under the
assumption $K+G-\mu-a\ne 0$ and $K-G-\lambda+b\ne 0$, $P(0)\ne
0$, $P(1)\ne 0$. Hence $0<\rho_1<\rho_2<1$ and $0<\rho<1$.
\end{remark}
\begin{remark}
Using the formulae
\begin{equation}
  1-k^2\sn^2(\xi)=\dn^2(\xi)=\frac{\rmd^2}{\rmd\xi^2}\ln\Theta(\xi)
  +\widetilde C
\end{equation}
where $\widetilde C$ is a certain constant, the previous solution
$\rho$ can be expressed by $\Theta$ function.
\end{remark}

In order to compute the $y$ and $t$-equations, we first write
down the expressions for $u$, $f$ and $g$. They are
\begin{equation}\fhl
   \begin{array}{l}
   \D u=\frac{2\I R_1R_2}{R_1^2-R_2^2}\E{\I(\alpha_1-\alpha_2)}\bigg(
    (K-\lambda-R_3^2\cos^2\theta_3)\cot\theta_1\\
   \D\qquad-R_3^2\cos\theta_3\sin\theta_3\E{\I\delta}\bigg)\\
   \D\quad=\frac{2\I R_1R_2}{R_1^2-R_2^2}\E{\I(\alpha_1-\alpha_2)}\bigg(
    (\mu+R_3^2\sin^2\theta_3-K)\tan\theta_1\\
   \D\qquad+R_3^2\cos\theta_3\sin\theta_3\E{-\I\delta}\bigg)\\
   \D|u|^2=\frac{4R_1^2R_2^2}{(R_1^2-R_2^2)^2}
    \bigg(K(R_3^2+\lambda+\mu-K)-\lambda\mu-\lambda R_3^2\sin^2\theta_3
    -\mu R_3^2\cos^2\theta_3\bigg)\\
   \D f=R_1R_3\E{\I(\alpha_1-\alpha_3)}\big(
    \cos\theta_3\cos\theta_1+\sin\theta_3\sin\theta_1\E{\I\delta}\big)\\
   \D g=R_2R_3\E{\I(\alpha_2-\alpha_3)}\big(
    -\cos\theta_3\sin\theta_1+\sin\theta_3\cos\theta_1\E{\I\delta}\big).
\end{array}\label{eq:ufgexp}
\end{equation}

With the help of MAPLE, (\ref{eq:newlpy}) and (\ref{eq:newlpt})
are reduced to the following simple equations:
\begin{equation}\fhl
   \begin{array}{l}
   \theta_{1,y}=\gamma_1\theta_{1,x}\qquad
   \theta_{3,y}=\gamma_1\theta_{3,x}\qquad
   \theta_{1,t}=\gamma_2\theta_{1,x}\qquad
   \theta_{3,t}=\gamma_2\theta_{3,x}\\
   \D\alpha_{1,y}=\gamma_1\alpha_{1,x}-\frac{2R_2^2K}{R_1^2-R_2^2}\qquad
   \alpha_{2,y}=\gamma_1\alpha_{2,x}
    -\frac{2R_1^2(R_3^2+\lambda+\mu-K)}{R_1^2-R_2^2}\\
   \D\alpha_{3,y}=\gamma_1\alpha_{3,x}-\frac{2R_2^2R_2^2}{R_1^2-R_2^2}\\
   \alpha_{1,t}=\gamma_2\alpha_{1,x}+C_{12}\qquad
   \alpha_{2,t}=\gamma_2\alpha_{2,x}-C_{21}\qquad
   \alpha_{3,t}=\gamma_2\alpha_{3,x}+C_3\\
   (\beta_j-\alpha_j)_y=\gamma_1(\beta_j-\alpha_j)_x\qquad
   (\beta_j-\alpha_j)_t=\gamma_2(\beta_j-\alpha_j)_x+C_0\quad(j=1,2,3)
\end{array}
\end{equation}
where the constants $\gamma_1$, $\gamma_2$, $C_0$, $C_{12}$, $C_{21}$ and
$C_0$ are given by
\begin{equation} \fhl
   \begin{array}{l}
   \D\gamma_1=\frac{R_1^2+R_2^2}{R_1^2-R_2^2}\\
   \D\gamma_2=\frac{2(R_1^4+R_2^4-(R_1^2+R_2^2)(R_3^2+\lambda+\mu))}{R_1^2-R_2^2}
   -\frac{4R_1^2R_2^2(R_3^2+\lambda+\mu-2K)}{(R_1^2-R_2^2)^2}
\end{array}
\end{equation}
\begin{equation}\fhl
   \begin{array}{l}
   \D C_0=\frac{2(\lambda-\mu)(R_1^4+R_2^4)}{R_1^2-R_2^2}\\
   \D C_{ij}=\frac 2{(R_1^2-R_2^2)^2}\Bigg(
    (R_j^4+2R_i^2R_j^2-R_i^4)\bigg((\lambda-\mu)(G-\frac{R_i^2-R_j^2}2)\\
   \D\qquad+(R_i^2-R_j^2)(\frac{\lambda+\mu}2-K_i)-\lambda\mu-\lambda R_3^2
    \bigg)+4R_1^2R_2^2K(R_3^2+\lambda+\mu-K)\\
   \D\qquad-\lambda(R_i^2-R_j^2)(R_1^4+R_2^4)
    -2R_j^4K_i^2\Bigg)\qquad(i,j)=(1,2)\hbox{ or }(2,1)\\
   K_1=K\qquad K_2=R_3^2+\lambda+\mu-K\\
   \D C_3=\frac 2{(R_1^2-R_2^2)^2}\Bigg(
    R_2^2(R_2^4+3R_1^4-R_1^2R_2^2)(\lambda+R_3^2)+
    R_1^2(R_1^4-R_2^4+R_1^2R_2^2)\mu\\
   \qquad -2(R_1^2+R_2^2)(R_1^4+R_2^4)K-2R_1^2R_2^2(R_1^4-R_2^4)\Big).
\end{array}
\end{equation}
Hence
\begin{equation}
   \rho=\rho_1+(\rho_2-\rho_1)\sn^2(p(x+\gamma_1 y+\gamma_2 t-x_0))
\end{equation}
where $x_0$ is an arbitrary constant, $p$ is given by
(\ref{eq:k-p}) and the parameter of the function sn is $k$
given by (\ref{eq:k-p}).

The solutions of the DSI equation are
\begin{equation}\fsl
   \begin{array}{rl}
   u=&\D\pm\frac{\I R_1R_2}{R_1^2-R_2^2}\;\frac{1}{\rho(\xi)\sqrt{1-\rho(\xi)}}\\
   &\D\times\bigg((2K-\lambda-\mu-a+b)\rho(\xi)+(\mu+a-K-G)
    -\I\sqrt{P(\rho(\xi))}\bigg)\\
   &\D\times\exp\bigg(\I\int Q(\rho(\xi))\,d\xi
    +\I\alpha(x-\gamma_1 y+\gamma_2 t)
    +\I(C_{12}+C_{21})t\bigg)
\end{array}\label{eq:uexp}
\end{equation}
and
\begin{equation}
   \begin{array}{l}
   v_1=2R_1^2((\mu-\lambda)\rho(\xi)+K-\mu)\\
   v_2=2R_2^2((\lambda-\mu)\rho(\xi)+K-\mu+a)
\end{array}
\end{equation}
where
\begin{equation}
   \begin{array}{l}
   \xi=x+\gamma_1 y+\gamma_2 t-x_0\\
   \D \alpha=\frac{R_2^2K-R_1^2(R_3^2+\lambda+\mu-K)}{R_1^2+R_2^2}\\
   \D Q(\rho)=\frac{2b\rho^2+(\mu-b-2G)\rho+(K+G-\mu-a)}{2\rho(1-\rho)}\\
   \rho(\xi)=\rho_1+(\rho_2-\rho_1)\sn^2(p\xi)
\end{array}
\end{equation}
and the parameter $k$ of the function sn is given by (\ref{eq:k-p}).

$u$ has no singularity when (\ref{eq:KGcond}) holds because in
this case $0<\rho<1$.

Suppose the minimal positive period of the function sn with
parameter $k$ is $T(k)$ and
\begin{equation}
   A=\frac{p}{T(k)}\int_0^{T(k)/p} Q(\rho(\xi))\,d\xi.
\end{equation}
Then we have the following properties of the solutions:

(1) $u$ is a double periodic function on $(x,y)$ plane. The
period for $x+\gamma_1 y$ is $T(k)/p$, while the period for
$(A+\alpha)x+(A-\alpha)\gamma_1 y$ is $2\pi$.

(2) $u$ is periodic with respect to $t$ if and only if
$\D\frac{2\pi p}{(C_{12}+C_{21}+A\gamma_2+\alpha\gamma_2)T(k)}$
is a rational number.

(3) $|u|^2$, $v_1$ and $v_2$ are periodic functions
of $x+\gamma_1 y+\gamma_2 t$ only, and they extend constantly in a
transversal direction on $(x,y)$ plane.

(4) The phase of $u$ depends not only on the linear functions of
$x$, $y$ and $t$, but also on an sn function of $x+\gamma_1 y+\gamma_2
t$. This can be obtained from (\ref{eq:uexp}) and
\begin{equation}
   (\arg u)_x=\re\frac 1\I\frac{u_x}{u}=\re\frac{2f\bar g}{\I u}
   \ne\hbox{constant}
\end{equation}
by using (\ref{eq:ufgexp}) and tedious computation.

It is still interesting to solve more general periodic solutions
using this method.

\section*{Acknowdgements}
This work was supported by the Chinese National Research Project
``Nonlinear Science'', the City University of Hong Kong (Grant No.\
7001041), the Research Grants Council of Hong Kong (Grant No.\
9040395, 9040466), the Doctoral Program Foundation and
the Key Project for Young Teachers of the Ministry of
Education of China, the National Natural Science Foundation of
China (Project 19801031) and the Special Grant of Excellent Ph.D.
Thesis. The authors thanks the referee for helpful suggestions.
The first author (Z.~X.~Zhou) is also grateful to the Department
of Mathematics of the City University of Hong Kong for the
hospitality.

\end{document}